\documentclass[preprint]{vgtc}               

\ifpdf
  \pdfoutput=1\relax                   
  \usepackage{graphicx}                
  \DeclareGraphicsExtensions{.pdf,.png,.jpg,.jpeg} 
\else
  \usepackage{graphicx}                
  \DeclareGraphicsExtensions{.eps}     
\fi%

\graphicspath{{figures/}{pictures/}{images/}{./}} 

\usepackage{microtype}                 
\PassOptionsToPackage{warn}{textcomp}  
\usepackage{textcomp}                  
\usepackage{mathptmx}                  
\usepackage{times}                     
\usepackage{cite}                      
\usepackage{tabu}                      
\usepackage{booktabs}                  
\usepackage{graphicx}
\usepackage{xpunctuate}
\usepackage{xspace}
\usepackage{enumitem}
\usepackage{comment}
\usepackage{svg}

\definecolor{forestgreen}{RGB}{34, 139, 34}
\newcommand{\revision}[1]{{\color{black}{#1}}}

\newcommand{\ie}{{i.e.,}\xspace}
\newcommand{\eg}{{e.g.,}\xspace}
\newcommand{\etal}{{et~al\xperiod}\xspace}
\newcommand{\aka}{{a.k.a.}\xspace}
\newcommand{\etc}{{etc\xperiod}\xspace}

\onlineid{0}

\vgtccategory{Research}

\vgtcinsertpkg

\ieeedoi{10.1109/VIS54172.2023.00028}

\title{WhaleVis: Visualizing the History of Commercial Whaling}

\author{
    Ameya Patil\thanks{e-mail: ameyap2@cs.washington.edu} \\ %
    \parbox{1.4in}{\scriptsize \centering Computer Science and Engineering \\ University of Washington, Seattle}
\and Zoe Rand\thanks{e-mail: zrand@uw.edu} \\%
    \parbox{1.4in}{\scriptsize \centering Quantitative Ecology and Resource Management \\ University of Washington, Seattle}
\and Trevor Branch\thanks{e-mail: tbranch@uw.edu} \\%
    \parbox{1.4in}{\scriptsize \centering School of Aquatic and Fishery Sciences \\ University of Washington, Seattle}
\and Leilani Battle\thanks{e-mail: leibatt@cs.washington.edu} \\%
    \parbox{1.4in}{\scriptsize \centering Computer Science and Engineering \\ University of Washington, Seattle}
}

\teaser{
  \centering
  \includegraphics[width=\linewidth]{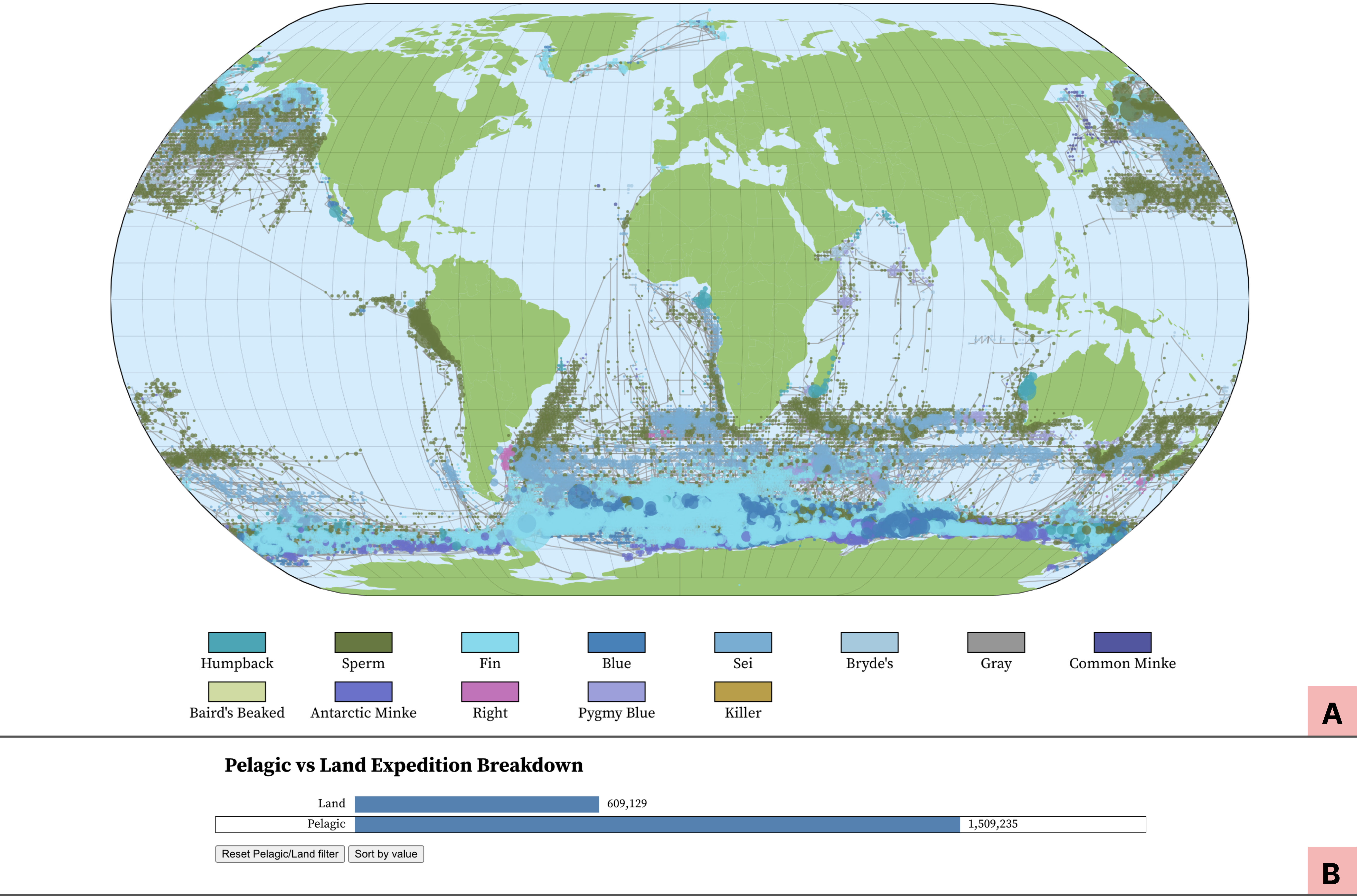}
  \caption{Selective visualizations from the WhaleVis dashboard: The map visualization \textbf{\textit{(A)}} shows pelagic (offshore) whale catches from 1880 to 1986 along with the routes traversed by whaling expeditions during this period. The bar chart \textbf{\textit{(B)}} shows the breakdown for pelagic vs land catches and also acts as an interface for setting a filter for pelagic whale catches. The route density in \textbf{\textit{(A)}} enables visual estimation of the search effort~\ie where whales were searched for. There are relatively fewer catches in the North Atlantic and South Pacific Oceans compared to other regions. Further, since fewer expeditions traversed those waters, we are aware of a relative reduction in search effort when inferring the whale populations from reported catches in those regions.}
  \label{fig:teaser}
}

\abstract {
    Whales are an important part of the oceanic ecosystem. Although historic commercial whale hunting~\aka whaling has severely threatened whale populations, whale researchers are looking at historical whaling data to inform current whale status and future conservation efforts. To facilitate this, we worked with experts in aquatic and fishery sciences to create WhaleVis—--an interactive dashboard for the commercial whaling dataset maintained by the International Whaling Commission (IWC). We characterize key analysis tasks among whale researchers for this database, most important of which is inferring spatial distribution of whale populations over time. In addition to facilitating analysis of whale catches based on the spatio-temporal attributes, we use whaling expedition details to plot the search routes of expeditions. We propose a model of the catch data as a graph, where nodes represent catch locations, and edges represent whaling expedition routes. \revision{This model facilitates visual estimation of whale search effort and in turn the spatial distribution of whale populations normalized by the search effort---a well known problem in fisheries research. It further opens up new avenues for graph analysis on the data, including more rigorous computation of spatial distribution of whales normalized by the search effort, and enabling new insight generation.} We demonstrate the use of our dashboard through a real life use case.
}

\CCScatlist{
  \CCScatTwelve{Applied computing}{Bioinformatics}{}{}; 
  \CCScatTwelve{Applied computing}{Environmental sciences}{}{} 
}


\begin{document}


\firstsection{Introduction}

\maketitle

Whales are critical to preserving our oceanic ecosystems~\cite{seilacher2005whale, flammang2020remoras, roman2014whales, heidi2022whales, pershing2010impact} and are thus important for ocean conservation efforts.
Effectively protecting whales requires understanding the spatial distribution of their populations. However, tracking whales is a high effort and low-yield task~\cite{hazen2017whalewatch, bamford2020comparison}. It requires scanning entire oceans for relatively small populations of whales, essentially searching for needles in a haystack. However, whale researchers observe that this is exactly what commercial whale hunting expeditions did to catch whales. Thus, commercial whale hunting~\aka whaling data could be used for altruistic purposes by using it to estimate changes in whale populations across the oceans over time.

However, commercial whaling data cannot be easily repurposed to infer the spatial distribution of whales. In particular, the whaling dataset emphasizes the \emph{results}~\ie catches, but not the \emph{effort} whalers placed in searching for whales, and where \emph{whales actually live}. Thus, there are multiple levels of inference that must be managed when translating whale catches (the knowns) into useful population data (the unknown). For example, whalers plan out a \emph{route} they will follow to search for whales; along this route, they may catch whales but they may also leave empty-handed. \emph{How do we take route information, whaling search effort, and total catches into account to better estimate whale populations across the oceans?}

In this paper, \revision{we take the first step towards addressing this question in the form of WhaleVis\footnote{\href{https://observablehq.com/@whales/whale-vis-dashboard-expedition-routes}{observablehq.com/whale-vis-dashboard-expedition-routes}},\footnote{ \href{https://observablehq.com/@whales/iwc-dataset-expedition-routes}{observablehq.com/iwc-dataset-expedition-routes}}---an interactive dashboard for analyzing millions of records of whaling data maintained by the International Whaling Commission (IWC).} WhaleVis was developed through a design study~\cite{sedlmair2012design, meyer2019criteria, ye2020user} in collaboration with domain experts in aquatic and fishery sciences.
The key idea of WhaleVis is to enable analysis of whale catches combined with that of whaling expedition routes, to estimate the spatial distributions of whale populations normalized by the search effort--- a well known and non-trivial problem in fisheries research involving many variables~\cite{skalski2005analysis}. Through WhaleVis, we transform the recorded whaling events, into a powerful tool for ocean conservation.




A major challenge for WhaleVis is the bias of the IWC data in favor of ``positive'' data points. It only records whales caught at a certain geographic location by a certain expedition, but does not record failed searches that occurred between locations, which are critical to understanding where whales actually were in the oceans. To address this limitation, we propose a graph representation of whaling data in WhaleVis, where we model the catch locations as nodes of the graph, and the expedition routes extracted from the data, as edges of the graph, refer~\autoref{fig:teaser}(A). 
\revision{In the current version of WhaleVis, the density of edges (expedition routes) serves as a visual proxy for search effort. This helps us to visually account for search effort while estimating whale population maps.
In the future, this graph representation can be leveraged to calculate search effort-normalized population maps rigorously using graph algorithms, e.g., network diffusion~\cite{guille2013information}.} It further opens up new avenues for performing graph analysis on the data which can uncover new insights. 




We develop WhaleVis in the Observable notebook environment, \revision{using DuckDB\cite{kohn2022duckdb-wasm} as the backend data processor, and D3.js for rendering.} We demonstrate the use of WhaleVis through a real life example in~\autoref{sec:use-case} showing how WhaleVis enables effective analysis of historical whaling events.

In summary, we make the following contributions:

\vspace{-\topsep}
\begin{itemize}[leftmargin=0.1in]
    \setlength\itemsep{0.01mm}
    
    \item characterization of the analysis tasks for the IWC whaling data,
    
    \item an interactive dashboard for understanding and generating insights from historical whaling events,
    
    \item a graph representation of the whaling data to \revision{facilitate visual estimation of whale search effort.}
\end{itemize}

\section{Related Work}



Relevant prior work in visualization dashboards for tracking wildlife consists of both generic wildlife mapping frameworks, and dedicated whale mapping dashboards. iNaturalist~\cite{inaturalist} is a citizen science tool focusing on data collection, with visualizations only to aid basic summarization of the data. EarthRanger~\cite{er2} on the other hand, is a feature rich interactive visualization and analysis dashboard which connects to multiple data sources. 
However, being too generic, both these tools do not cater to the kind of analysis whale domain experts are interested in.

Dedicated whale mapping dashboards include WhaleMap \cite{whalemap}, HappyWhale \cite{happywhale}, Pacific Whale \cite{pacificwhale}, Obis-Seamap \cite{obis-seamap} and WhaleTest \cite{whaletest}. All these dashboards map whale sightings, but not whaling events. Although Obis-Seamap and WhaleMap can connect to multiple data sources, they lack the level of transparency to easily incorporate the IWC whaling data. These interfaces also suffer from user experience challenges (1) sub-optimal visualization designs hindering interpretability of the data—Obis-Seamap, WhaleMap; (2) unintuitive interactions for querying the data—HappyWhale, Obis-Seamap, PacificWhale, WhaleTest; and (3) high response time for interactions (tens of seconds) to query the full data—HappyWhale, Obis-Seamap, WhaleMap.


To facilitate spatial analysis based on geographic coordinates, we stick to the standard visualization approach for wildlife mapping~\ie scatter-plots on geographic maps~\cite{obis-seamap, happywhale, whalemap, whaletest, pacificwhale, er2, inaturalist}. Scalable scatter-plots are employed for large datasets to avoid occlusion of marks and ease cognition~\cite{guo2018efficient, tao2020kyrixs}. These work by aggregating data to show fewer visual marks and visualizing details on demand. This approach is used in HappyWhale and Obis-Seamap. We follow the same idea although with a slight change to better facilitate whale researchers, refer~\autoref{sec:dnd}. We also draw upon best practices in dashboard  and interaction design~\cite{sarikaya2019dashboards, bach2022dashboard, lam2008framework} to facilitate intuitive interactions and effective analysis experience.

We develop WhaleVis with domain experts, while taking the best from both, the interaction rich nature of EarthRanger, and the dedicated whale mapping dashboards.


\section{Dataset}

We use the whaling data originally collated by the Bureau of International Whaling Statistics (BIWS) and currently maintained by the International Whaling Commission (IWC)~\cite{allison2020iwci, allison2020iwcs}. The current version of the dataset (V7.1) covers whaling events from 1880 to 2020\footnote{Data from 1986 onwards currently unavailable in WhaleVis awaiting IWC approval} across all the oceans. It has details about whale catches like date, geographic coordinates, species, length, sex of the whales~\etc, and about the corresponding expedition which caught the whale like the expedition code, nation, company name, expedition type~\etc. The data is available as a set of .csv files which were merged together. Records with missing timestamp and geographic coordinates were removed to avoid noise. We thus removed \revision{1.4\% of all the whale catch records}, leaving a total of 2,148,279 records.

Originally collected for commercial book-keeping, whale researchers today are interested in using this data for whale conservation and management. This requires considerable time and effort to transform the data, for repurposing it. In the next section, we describe the kind of questions whale researchers ask of this dataset, and show how WhaleVis facilitates answering these questions effectively. In~\autoref{sec:use-case}, we show how WhaleVis is able to quickly replicate an analysis which otherwise costed significant time and effort for whale researchers.

\section{Design Study \& Implementation Details}
\label{sec:dnd}

We worked with our coauthors in aquatic and fishery sciences over four months, following best practices for design studies~\cite{sedlmair2012design, meyer2019criteria, ye2020user}. Here, we describe the analysis themes and constituent concrete analysis tasks gathered from them.

\subsection{Analysis Themes \& Tasks}
\label{sec:atnt}



\begin{enumerate}[label={}, leftmargin=0in, nosep]
    \item \textbf{1. Understanding spatio-temporal distribution of whale catches.} These tasks are about seeing the big picture of how humans hunted  whales across space and time, with possible correlations with physical attributes of the whales caught.

    \begin{itemize}[label={}, leftmargin=0.15in]
        \item \textbf{\textit{(T1)}} Understand how whaling spread from one region to another over time.
        \item \textbf{\textit{(T2)}} Identify areas where a certain whale species was caught in the past, but recent surveys do not report any sightings, suggesting possible extirpation.
        \item \textbf{\textit{(T3)}} Identify examples of vagrant whales: catches reported far away from clusters of other catches and the known range of existing species populations.
        \item \textbf{\textit{(T4)}} Inspect the characteristics of the whales caught in a given region and time period (species, sex,  length,~\etc) or by a particular whaling expedition or country.
    \end{itemize}

    \smallskip
    \item \textbf{2. Inferring spatial distribution of whale populations from whaling events.} These tasks focus on generating insights about the ground truth whale population numbers from the whaling events. However whaling data only tells us where whales were caught. Accordingly, the following constituent tasks show the steps involved in understanding the ground truth distribution from the whaling events.

    \begin{itemize}[label={}, leftmargin=0.15in]
        \item \textbf{\textit{(T5)}} Map the effort spent by whaling expeditions over space and time~\ie search effort.
        \item \textbf{\textit{(T6)}} Compute spatial distribution of catches over time, normalized by the search effort.
    \end{itemize}

    Note that the current state of the catch database does not make the search effort data readily available.

    \smallskip
    \item \textbf{3. Inform management and conservation efforts.} While the first two analysis themes were concerned about the history of whaling, this theme targets the present and future of whales.

    \begin{itemize}[label={}, leftmargin=0.15in]
        \item \textbf{\textit{(T7)}} Extract data subsets pertaining to certain species, sex or region, for use in models for assessing current population status.
    \end{itemize}
\end{enumerate}

\noindent These tasks are not exhaustive but represent the kind of questions domains experts ask of this dataset.









\subsection{Design Goals \& Implementation}
Based on the analysis themes and tasks, best practices in interactive dashboard design, and future scope of use of our tool, we chose three design goals. We now describe these design goals and share corresponding implementation details.

\smallskip
\noindent \textbf{G1: Enable Spatial \& Graph Analysis.} The first two analysis themes are concerned with the spatial analysis of whaling events which can be performed using the spatio-temporal data attributes. However, tasks in the second theme cannot be performed directly using those attributes. We recognised that mapping the search effort \textbf{\textit{(T5)}} and computing catches normalized by search effort \textbf{\textit{(T6)}} could be better facilitated if we model the catch data as a graph---nodes representing catch locations, and edges representing expedition routes between catch locations. \revision{This approach contributes (1) a visual estimation of the search effort by using the route density as a visual proxy, and (2) a framework through which graph analysis algorithms can be applied to compute the search effort as a continuous distribution along the expedition routes, as opposed to a discrete distribution only at the catch locations.
We aim to implement graph analysis algorithms for this task as future work. In this way, we can support seamless transitions between spatio-temporal analysis and visual graph analysis in future versions of WhaleVis.}

\noindent \textbf{Implementation.} We visualize scatterplot and binned heatmap of catches on a geographic map to enable spatial analysis, with scalable scatterplot styled multiple levels of aggregation to handle possible overplotting~\cite{tao2020kyrixs}. We allow setting the aggregation level independently of the zoom level to view global level spatial trends in whaling events, which is more helpful compared to viewing local trends in a small zoomed-in region. \revision{We set default semantic color encoding for the map visualization based on the scientific classification taxonomy for whale species~\cite{cetaxonomy}, continental classification of nations, and globally recognised color schemes for sex and land vs. pelagic catches. We plan to support user configurable color encoding for better accessibility support, as future work.}

\begin{figure*}[t]
    \centering
    \includegraphics[width=\textwidth]{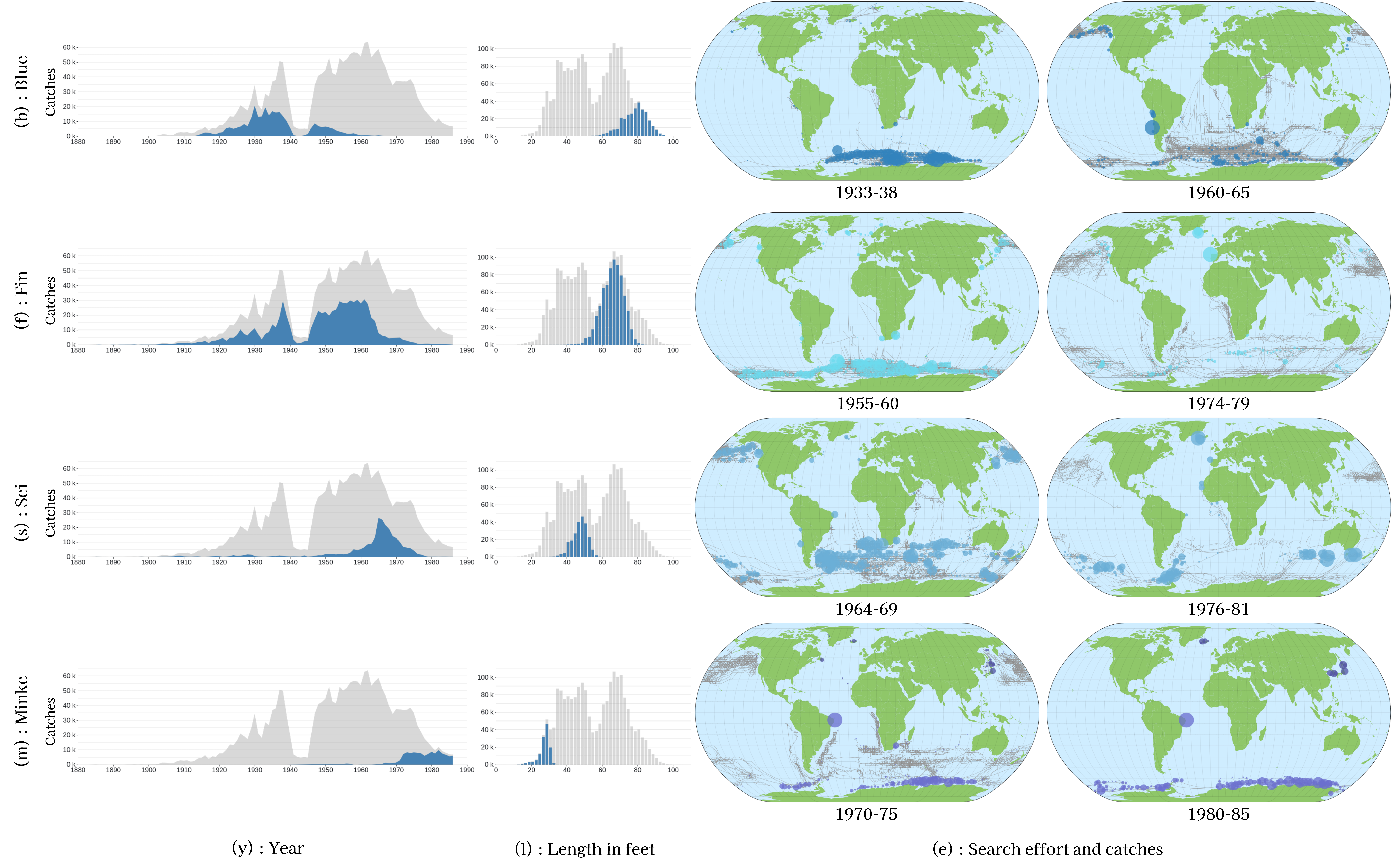}
    \caption{Progression of whaling species over time: blue $\rightarrow$ fin $\rightarrow$ sei $\rightarrow$ minke. Columns (y) and (l) show the timeline and the distribution of lengths respectively for each species. Column (e) shows the expedition routes (gray lines) and catches (circles) during the periods of peak whaling (3rd column) and decline in whaling (4th column), for each species.}
    \label{fig:progression}
    \vspace{-5mm}
\end{figure*}

Additional pre-processing is performed to model the data as network graph. We reconstruct the routes for each expedition from the locations of its successive catches; a route is created whenever the location over successive catches changes, indicating movement of the expedition. We extracted 97,713 such routes (edges). These expedition routes are also visualized on the geographical map, refer~\autoref{fig:teaser}(A) and~\autoref{fig:progression}(e).

\smallskip
\noindent \textbf{G2: Balance Inclusivity \& Usability.} The ability to support as many different kinds of analyses as possible (inclusivity), was a recurring theme in our design discussions. Each such use case entailed a new visualization or encoding, and corresponding UI widgets to set the encoding or filter data as needed. However, prior work~\cite{lam2008framework} has shown that too many visual or interactive elements can decrease usability. Users generally prefer having fewer UI elements, and consistent and easy to use interactions. We strive to reduce such visual clutter and balance inclusivity and usability, by adhering to best practices of dashboard design~\cite{sarikaya2019dashboards, bach2022dashboard}.


\noindent \textbf{Implementation.} We prioritized a default set of visualizations in WhaleVis through our design discussions. More precisely, we chose the data attributes covered in the analysis tasks \textbf{\textit{(T1-T7)}} as a reasonable starting point for whale researchers to meet their current analysis requirements. Currently, WhaleVis visualizes catch attributes: lat, lon, date, species, sex and length; and expedition attributes: nation, expedition type, and the expedition routes. Supporting configurable visualizations as future work is the motivation behind our next design goal \textbf{G3}. To keep UI elements to a minimum, we cross-link all the visualizations and enable interactions through direct manipulations~\cite{hutchins1985direct}. All the visualizations were implemented using D3.js.

\smallskip
\noindent \textbf{G3: Transparency \& Sustainability.} The IWC catch dataset was meticulously digitized from multiple hand-written log books over decades~\cite{allison2020iwci, allison2020iwcs} and continues to be updated even today through community involvement.
In the same spirit, we aim to build a tool which keeps the data transformation and visualization pipeline transparent. This would reduce the barrier for the whale researchers to easily explore/analyze the data, participate in improving the tool with additional features to be implemented and insights to be shared~\cite{isenberg2011collaborative}, and suggest corrections for possibly biased data processing/visualization~\cite{correll2019ethical}. This becomes critical also considering the growth of whale monitoring data, and sustaining the tool in terms of additional datasets, use cases and scalability. 

\noindent \textbf{Implementation.} We use the Observable notebook environment to leverage its transparency and reactive development framework. We create separate notebooks for data transformation and the dashboard implementation to keep the implementation modular and facilitate reuse. We also facilitate downloading filtered subsets of the data for offline analysis\footnote{Currently disabled in the dashboard awaiting approval from IWC}. We use DuckDB-Wasm~\cite{kohn2022duckdb-wasm} both for the data transformation and as our data store.



\section{Use case - Progression of Whaling Species}
\label{sec:use-case}

We refined WhaleVis based on the feedback from managers of the IWC whaling dataset, in addition to the guidance from our domain expert coauthors. We presented this work in the IWC Scientific Committee Meeting~\cite{patil2023whalevis} and are in the process of conducting a formal user study for further improvements. As a use case, we replicate and extend an analysis originally performed by Hilborn~\etal~\cite{hilborn2003state} which took them days to perform, but was performed within seconds by one of our domain expert coauthors. 

\autoref{fig:progression} shows us how modern commercial whaling depleted one species before moving on to the next, depending on the abundance, ease of hunting, and the commercial gain (amount of oil) that could be obtained from each species. Blue whales (b) followed by fin whales (f) were two of the most heavily hunted species at the start of the 20th century. Blue whaling declined during the 1960s, due primarily to a decline in their numbers as illustrated by the declining catches despite increased search effort over time (b,e). By 1970, as fin whaling started declining for possibly similar reasons (f,e), sei whaling peaked (s). The decline of sei whaling during the 1970-80, overlapped with a peak in minke whaling (m). The length histograms (l) show one possible reason behind such a progression: blue whales are the longest (80 ft.) and thus yield high commercial gains, compared to fin whales (70 ft.), followed by sei (50 ft.) and minke whales (30 ft.).

Declining population was not as much a reason for decline in minke whaling as it was for blue, fin and sei whaling (m,e). Whaling activities in general reduced drastically after the IWC issued a ban on commercial whaling starting 1986.

\section{Future work \& Conclusion}

Currently WhaleVis only visualizes the search \emph{paths} of whaling expeditions. We envision supporting graph analysis methods for rigorous analysis of the search \emph{effort}. An immediate use is to compute realistic estimates of catches normalized by search effort, and thus the spatial distribution of whale populations, interpolated as a continuous function along the search path. We also plan to generalize such graph visualization and analysis constructs for route analysis,~\eg network based diffusion~\cite{guille2013information} for multiple domains like tracking animal movements, atmospheric/oceanic currents,~\etc

The IWC whaling data is only as valuable as the insights that can be gleaned from it, making the domain knowledge of whale researchers just as important as the data itself. Although there are established techniques for analyzing large tabular datasets, which we employ in WhaleVis, we lack commensurate methods for managing the \emph{insight corpora}. In future research, we seek to explore how might whole communities externalize, validate, and share insights with not just each other, but also the non-experts to increase awareness about critical global problems like conservation of wildlife~\cite{ynnerman2016exploranation}.


The importance of historical whaling data in understanding the current whale population distributions cannot be undermined. However to make WhaleVis a complete tool, we also plan to incorporate interactive ``what-if" modeling scenarios~\cite{gathani2022augmenting} to enable making predictions to better support future conservation efforts. 

We hope that WhaleVis can thus streamline the process of gathering data about and understanding, whale population distributions, and also be used to effectively organize management actions for the conservation of whales.


\acknowledgments{
    This project was funded in part by the University of Washington Computing for the Environment Initiative and NSF Award IIS-2141506.
}

\bibliographystyle{abbrv-doi}

\bibliography{template}
\end{document}